\documentclass[preprint,aps,prd,nofootinbib]{revtex4-1}
\usepackage{graphicx,makecell} 
\usepackage{epstopdf}
\usepackage{amsmath}
\usepackage{color}
\usepackage{subcaption}
\usepackage{mathrsfs}
\usepackage{hyperref}

\usepackage{amsfonts}
\usepackage{amssymb}
\usepackage{bm}

\def\bea{\begin{eqnarray}}
\def\eea{\end{eqnarray}}
\def\<{\langle}
\def\>{\rangle}

\def\uaua{{\uparrow\uparrow}}
\def\uada{{\uparrow\downarrow}}
\def\daua{{\downarrow\uparrow}}
\def\dada{{\downarrow\downarrow}}
\def\be{\begin{equation}}
\def\ee{\end{equation}}

\def\non{\nonumber}

\def\amp{{\mathcal{A}}}
\def\ss{{\text{s}}}

\def\fun{{\mathcal{F}}}
\def\flm{{\mathcal{G}}}

\begin{document}

\title{Quantum  Magic in Quantum Electrodynamics}

\author{\vspace{0.5cm}Qiaofeng Liu$^a$, Ian Low$^{a,b}$, and Zhewei Yin$^{a,b}$}
\affiliation{\vspace{0.5cm}
\mbox{$^a$Department of Physics and Astronomy, Northwestern University, Evanston, IL 60208, USA}\\ 
\mbox{$^b$High Energy Physics Division, Argonne National Laboratory, Lemont, IL 60439, USA}\\
}

\begin{abstract}
In quantum computing, non-stabilizerness -- the magic -- refers to the computational advantage of certain quantum states over classical computers and is an essential ingredient for universal quantum computation. Employing the second order stabilizer R\'enyi entropy to quantify magic, we  study the production of  magic states in Quantum Electrodynamics (QED) via  2-to-2 scattering processes involving electrons and muons. 
 Considering all 60 stabilizer initial states, which have zero magic, the angular dependence of  magic produced in the final states is governed by only a few patterns, both in the non-relativistic and the ultra-relativistic limits. Some processes, such as the low-energy $e^-\mu^-\to e^-\mu^-$ and Bhabha scattering $e^-e^+\to e^-e^+$, do not generate magic at all.  In most cases the largest magic generated is significantly less than the maximal possible value of  $\log (16/7) \approx 0.827$. The only instance where QED is able to generate maximal magic is the low-energy  $\mu^-\mu^+\to e^-e^+$, in the limit $m_e/m_\mu \to 0$, which is well approximated in nature. Our results suggest QED, although  capable of producing maximally entangled states easily, may not be an efficient mechanism for generating quantum advantages.

\end{abstract}

\maketitle


\tableofcontents

\section{Introduction}

Quantum computing and quantum simulations hold great promise for the potential to solve complex problems and simulate intricate systems which are intractable for classical computers. In particular, entanglement as the most prominent feature of quantum mechanics has long been cited as the primary ``quantum resource'' for quantum computers. Indeed, entanglement plays a central role in  the celebrated Shor's algorithm \cite{Shor:1994jg} for efficient factorization of large numbers into prime numbers.

Despite its fundamental role in quantum mechanics, entanglement alone is not sufficient to achieve quantum advantage over classical computers. This is demonstrated by the Gottesman-Knill theorem \cite{gottesman1998heisenbergrepresentationquantumcomputers,Gottesman:1999tea,Aaronson:2004xuh}, which states that certain commonly employed quantum circuits,\footnote{These are circuits involving only the so-called Clifford gates: Hadamard (H), Controlled NOT (CNOT) and Phase (S) gates.} even if involving highly entangled  states, can be simulated efficiently using classical computers. The theorem suggests that new ingredients, other than entanglement, are needed to identify the true computational power of quantum realm. 

To address the limitation from the Gottesman-Knill theorem, Bravyi and Kitaev introduced the notion of a magic state in Ref.~\cite{Bravyi:2004isx} and showed that it is an essential ingredient for universal quantum computation. Magic is a new quantum resource which represents some intrinsic ``quantumness'' that cannot be simulated efficiently using classical algorithms. It is then  an important task to characterize this class of states \cite{Howard:2017maw,Beverland:2019jej,Wang:2020nyx} and quantify the presence of magic \cite{Mari:2012ypq,Veitch_2012,Emerson:2013zse,Bravyi_2019,Leone:2021rzd,Warmuz:2024cft}. Among the  measures of magic the stabilizer R\'enyi entropy (SRE) proposed in Ref.~\cite{Leone:2021rzd}, which is based on the R\'enyi entropy of order $\alpha$,  has attracted considerable attention due to its ease to compute and wide-ranging applications \cite{Oliviero:2022bqm,Wang:2023uog,Haug:2023ffp,Lami:2023naw,Tarabunga:2024ugl,Leone:2024lfr}. 

Another interesting aspect is to identify states which are most distinct from the zero magic states, i.e. quantum states with maximal magic and, therefore, maximal quantum resources. It was argued in Refs.~\cite{Leone:2021rzd,Cuffaro:2024wet} that, for a $d$-dimensional Hilbert space, an upper bound for the second order SRE is $\log \frac{d+1}2$. For a two-qubit system, $d=4$ and the upper bound is $\log(5/2) \approx 0.916$. However, Ref.~\cite{Cuffaro:2024wet} pointed out that the bound can only be saturated by a single qudit but not by a composite two-qubit system. Very recently we derived a more stringent bound of  $\log({16}/{7})\approx 0.827$ for two-qubit states and also obtained all 480 states saturating the maximum \cite{Liu:2025frx}.  

The study of magic has progressed beyond quantum algorithms and appeared in diverse systems, including quantum phases of matter \cite{Ellison:2020dkj}, conformal field theories \cite{White:2020zoz}, black hole information \cite{Hayden:2007cs,Leone:2022afi}, nuclear forces \cite{Robin:2024bdz}, dense neutrino systems \cite{Chernyshev:2024pqy},  massless scalar field theories \cite{Nystrom:2024oeq,Cepollaro:2024sod},  quantum many-body systems \cite{Oliviero:2022euv,Rattacaso:2023kzm,Gu:2024ure}, simulations of quantum gravity \cite{Cepollaro:2024qln}, as well as top quark productions at the Large Hadron Collider \cite{White:2024nuc}.

Given the central importance of magic in universal quantum computation and designing fault-tolerant quantum algorithms, understanding what  processes can create and destroy magic has become an important topic \cite{Niroula:2023meg}. In this work we initiate a study on the production of two-qubit magic states in QED, which governs the interaction of light and matter and is viewed as one of the most important theories in physics. That is we consider an initial two-qubit state with zero magic, which interact via QED, and compute the magic of the outgoing qubits. More specifically, we would like to examine the ability of QED to generate quantum resource -- the magic -- in 2-to-2 scattering processes involving electrons and muons. The subject of particle scattering and decay in an information-theoretic context is a growing area of research and, with the exception of Ref.~\cite{White:2024nuc}, most works focus on entanglement related quantities \cite{ATLAS:2023fsd,CMS:2024pts,Peschanski:2016hgk,Fan:2017mth,Cervera-Lierta:2017tdt,Beane:2018oxh,Low:2021ufv,Liu:2022grf,Carena:2023vjc,Cheung:2023hkq,Liu:2023bnr,Hu:2024hex,Aoude:2024xpx,Kowalska:2024kbs,Low:2024mrk,Low:2024hvn,Horodecki:2025tpn}. Moreover, it is worth highlighting that QED is capable of generating maximally entangled states in scattering processes, as shown in Ref.~\cite{Cervera-Lierta:2017tdt}. 
Given that magic is the measure of quantum advantage, we would like to extend the study of quantumness in scattering processes beyond entanglement. In particular, using QED as a starting point, we seek to answer if any of the fundamental interactions in nature is an efficient machinery at generating quantum resources.

This paper is organized as follows. In the next Section we briefly review the concept of magic and the definition of SRE employed through this work. Then in Section \ref{sect:3} we compute the scattering amplitudes for QED processes involving electrons and muons in the low-energy limit and present angular distributions of final state magic for all 60 stabilizer initial states. The high-energy limit and the associated distributions of magic are given in Section \ref{sect:4}. In the last Section we present the discussion. For completeness we list all 60 stabilizer states for two-qubit systems in Appendix \ref{app:ss2q}.

\section{Magic in Quantum Computing}

This section provides a brief overview  of magic in quantum information. For more extensive discussions we refer to \cite{Nielsen:2012yss,Leone:2021rzd}. We use the following notation to represent the Pauli matrices: $I=\sigma^0$, $X=\sigma^1$, $Y=\sigma^2$ and $Z=\sigma^3$. The Pauli group $G_n$ for $n$-qubit is defined as
\begin{equation}
    G_n = \left\{ \phi\, P_1 \otimes P_2\otimes\dots\otimes P_n\  |\   P_i \in 
    \left\{ I, X, Y, Z \right\}\ {\rm and}\ \phi \in \{\pm 1, \pm i\}\ \right\}\ ,
\end{equation}
where $P_i$ acts on the $i$-th qubit. The phase $\phi$ is needed so that the Pauli strings  $P_1 \otimes P_2\otimes\dots\otimes P_n$ form a group under multiplication. As a result, the number of elements in  $G_n$ is $|G_n|=4^{n+1}$. Because of properties of Pauli matrices, any two elements of $G_n$ either commute or anticommute and  every element squares to $\pm I$.

A  stabilizer state $|\psi\rangle$ is  an eigenstate of some elements of $G_n$ with $+1$ eigenvalue:
\be
g |\psi\rangle = |\psi\rangle \ , \qquad g \in G_n \ .
\ee
The set of such elements  forms an abelian subgroup of $G_n$ called the stabilizer group. For every stabilizer state there is a maximal stabilizer group $S$ that is uniquely associated with $|\psi\>$. The stabilizer group $S$ has $2^n$ elements but only  $n$ generators whose products generate the entire group.\footnote{More generally, a finite group $G$ has at most $\log_2|G|$ independent generators.} Moreover, $-I$ doesn't stabilize any state and can't be an element of $S$. Given that elements of $G_n$ square to $\pm I$, this implies $g^2=I$ for all $g\in S$.  It was shown in Ref.~\cite{Aaronson:2004xuh} that the number of $n$-qubit pure stabilizer states is 
\be
N=2^n \prod_{k=0}^{n-1} \left(2^{n-k}+1\right) \ .
\ee
For $n=1$ there are six stabilizer states, corresponding to eigenstates of $X, Y$ and $Z$. For two-qubit there are 60 stabilizer states, which are listed in Refs.~\cite{garca2017geometrystabilizerstates,Robin:2024bdz}, many of which are highly entangled. 

The fact that $S$ is uniquely associated with a stabilizer state $|\psi\>$ and that it only has $n$ generators points to the computational advantage of using the stabilizer group to describe the dynamics of $|\psi\>$. Consider a unitary operation $U$ on a stabilizer state $|\psi\rangle$. For any element $g$ of $S$,
\be
U|\psi\rangle = U g|\psi\rangle = UgU^{\dagger}\, U|\psi\rangle \ ,
\ee
which demonstrates that $U|\psi\rangle$ is stabilized by $USU^\dagger$. Instead of using $2^n$ amplitudes to specify $U|\psi\rangle$,  one can simply specify the $n$ generators in $USU^\dagger$.

The stabilizer formalism is especially powerful when we focus on the set of unitary operations which preserves the Pauli group upon conjugation. These are the normalizers of the Pauli group: $N(G_n) = \{ U\ |\ UG_n U^\dagger = G_n\}$. When a quantum circuit involves only stabilizer states and unitary operations in $N(G_n)$, the dynamics of the states can be easily followed using the stabilizer formalism, since $USU^\dagger$ stays in $G_n$ for $U\in N(G_n)$. The normalizer group of $G_n$ is called the Clifford group and are   generated by the Hadamard (H), the CNOT and the Phase (S) gates, which are called the Clifford gates.

Gottesman-Knill theorem \cite{gottesman1998heisenbergrepresentationquantumcomputers,Gottesman:1999tea,Aaronson:2004xuh} then states that a quantum computation can be simulated efficiently, in polynomial time, on a classical computer if 1) the states are prepared in the computational basis, 2) the operation is implemented using only  the Clifford gates,  3) the measurement only involves the Pauli gates and 4) the output contains classical manipulations.  Under these circumstances quantum circuits, even if involving highly entangled states,  do not possess computational advantages over classical algorithms. 

The Clifford circuits appear in  many quantum protocols such as teleportation and entanglement purification. Repeated applications of Clifford gates to a $n$-qubit product state, say $|0\>^{\otimes n}$, will eventually produce all stabilizer states. That is, the stabilizer states constitute the orbit of the Clifford group. However, the Clifford circuits alone do not achieve universal quantum computation and a non-Clifford gate, such as the T-gate, needs to be added to achieve universality. In other words, repeated applications of Clifford and T gates will provide access to any quantum circuits.

The inclusion of non-Clifford gates represents the true quantum resources which cannot be obtained classically. In Ref.~\cite{Bravyi:2004isx} Bravyi and Kitaev showed that, the effect of introducing the T-gate to the quantum circuit can be equivalently described by using non-stabilizer states in Clifford circuits. Such non-stabilizer states are dubbed ``magic'' states due to their computational advantage over classical computers. Quantifying the new quantum resource represented by the magic then becomes an important question.

Ref.~\cite{Leone:2021rzd} introduces a measure called stabilizer R\'enyi entropy (SRE) of  order $\alpha$, for an integer $\alpha\ge 2$, which quantifies the non-stabilizerness - the \textit{magic} - of any pure state. Let $\mathcal{P}_n$ be the set of $n$-qubit Pauli strings with phase $+1$,
\begin{equation}
    \mathcal{P}_n = \left\{P_1 \otimes P_2\otimes\dots\otimes P_n \right\} , \quad  P_i \in 
    \left\{ I, X, Y, Z \right\} \ ,
\end{equation}
Introducing the density matrix $\rho = |\psi\>\<\psi|$ and
\be
\Xi_P(| \psi \>) =\frac1d  \left\langle\psi\vert P \vert \psi \right\rangle^{2} =\frac1d {\rm Tr}(\rho P)^2\ ,
\ee
the stabilizer $\alpha$-R\'enyi entropy of a state $\left|\psi\right\rangle$ is defined as
\bea
    M_\alpha \left(\left|\psi\right\rangle\right) &=& \frac{1}{1-\alpha} \log\, \sum_{P\in \mathcal{P}_n} \Xi_P^\alpha(| \psi \>)- \log d \nonumber \\
    &=& \frac{1}{1-\alpha} \log\, \sum_{P\in \mathcal{P}_n} \frac1d  \left\langle\psi\vert P \vert \psi \right\rangle^{2\alpha} \ .
\eea
where $d = 2^n$. As pointed out in Ref.~\cite{Leone:2021rzd}, $\sum_{P\in \mathcal{P}_n} \Xi_P(| \psi \>)= \left({\rm Tr} |\psi\>\<\psi|\right)^2=1$ and $\Xi (| \psi \>)\ge 0$. Therefore $\Xi_P(| \psi \>)$ can be viewed as the probability of finding $P$ in the density matrix $\rho=|\psi\>\<\psi|$. For a stabilizer state $\Xi_P(| \psi \>)=\pm 1$ for $d$ mutually commuting Pauli strings and vanishes otherwise. Therefore $M_\alpha$ vanishes for a stabilizer state. The SRE is also invariant under Clifford operations: $|\psi\>\to C|\psi\>$, where $C\in\{{\rm H}, {\rm CNOT}, {\rm S}\}$. For a product state the SRE is additive: $M_\alpha \left(\left|\psi\right\rangle\otimes|\phi\>\right) = M_\alpha \left(\left|\psi\right\rangle\right)+M_\alpha \left(|\phi\>\right)$. For  magic states, $M_\alpha> 0$.

In the following, we will choose to use
\bea
M_2 (| \psi \>) = - \log \Xi_2(|\psi\>) \ , \qquad \Xi_2(|\psi\>)\equiv \sum_{P\in \mathcal{P}_n} \frac{\left\langle\psi\vert P \vert \psi \right\rangle^{4}}4
\eea
as our measure of magic in 2-qubit states and present analytic expressions of $\Xi_2$ in our results.

Of great interest are states with maximal magic. Ref.~\cite{wang2023stabilizer,Cuffaro:2024wet} presented an upper bound of SRE,
\be
\label{eq:srebound}
M_\alpha (| \psi \>) \le \frac{1}{1-\alpha}\log \frac{1+(d-1)(d+1)^{1-\alpha}}{d} \ ,
\ee
which is saturated if and only if $|\psi\>$ belong is a Weyl-Heisenberg covariant Symmetric Informationally Complete state (WH SICs), when such a state exists in dimension $d$. Whether such states exist in all dimensions is a long-standing open problem \cite{axioms6030021} and turns out to be related to  Hilbert's 12th problem in algebraic number theory \cite{Appleby_2017}. For two-qubit states $d=4$ and the bound  in Eq.~(\ref{eq:srebound}) is never saturated \cite{Cuffaro:2024wet}. 

Recently in  Ref.~\cite{Liu:2025frx} we showed the maximum of $M_2$ for two-qubit states is  
\be
M_2 \le \log\frac{16}{7}\approx 0.827\ , 
\ee
and present all  two-qubit states satisfying the above bound, which turn out to be the minimally unbiased bases (MUB's) formed by the orbit of the WH group. For comparison, the bound obtained from Eq.~(\ref{eq:srebound}) is  $\log (5/2)\approx 0.916$, which is less stringent. We also conjecture that, when WH-SICs do not exist, the WH-MUBs are the maximal magic states and have the following SRE:  
\bea
M_\alpha (|\psi\>) = \frac{1}{1-\alpha} \log \frac{1 + (d-1) d^{1-\alpha}}{d},\label{eq:srewhmub}
\eea
which is clearly more stringent than the bound given in Eq.~(\ref{eq:srebound}). For two-qubit, $d=4$, and $\alpha=2$, the above gives $\log 16/7$. For $n$-qubit states, WH SICs exist only for $n=1$ and $n=3$ \cite{GODSIL2009246}. We conjecture that the bound in Eq.~(\ref{eq:srewhmub}) applies to $n$-qubit for $n\neq 1$ and $3$.

\section{Magic in QED at Low Energies}
\label{sect:3}

We will compute  scattering processes of muons and electrons in QED. The two particles are viewed as two qubits in the spin space, as their spin projection along a particular axis carry two possible values: $\pm 1/2$.  Since the spin projections in general are frame-dependent, it is essential to have a clear, consistent choice of reference frame. In other words, expectation values of the Pauli string operators are not rotationally invariant and depend on the choice of basis. Our choice is the spin-projection along the incoming beam axis in the centre-of-mass (CM) frame, with particle 1 traveling in the $+z$ direction and particle-2 in the $-z$-direction. Let $\vec{p}_{1,2}$ and $\vec{k}_{1,2}$ be the initial and final state 3-momenta in the CM frame, our construction of reference frame is given as follows:
\bea
\hat{z} = \hat{p}_1, \qquad \hat{y} = \frac{\vec{k}_1 \times \vec{p}_1}{|\vec{k}_1 \times \vec{p}_1|}, \qquad \hat{x} = \hat{z} \times \hat{y}.
\eea
This is a common choice of coordinate system in high energy collider physics. (See, for example, Refs.~\cite{Gao:2010qx,Gainer:2011xz}.)
The orientation of the $x$ and $y$ axes is  such that, for the final states, the azimuthal angle $\phi$ is given by $\phi=0$ for particle 1 and $\phi= \pi$ for particle 2. For instance, the 2-qubit state $|\uada\>$ means that the spin of particle 1 is in the $+z$-direction, while the spin of particle 2 is in the $-z$-direction. 
The amplitudes and magic computed in this frame will thus only depend on $\theta$ but not $\phi$.

It is worth emphaiszing that we will not be adopting a helicity basis, which projects the spin along the direction of motion, and insist on spin projections to the $z$-axis for both incoming and outgoing particles. The reason being the computation of magic depends on the choice of axis and a meaningful comparison  before and after the collision relies on a common axis to project the spin.

We  calculate the scattering amplitude ${\cal A}$ for a given process $i\to f$,
\be
S = 1+ i T \ ; \qquad \<f| T |i\> = (2\pi)^4 \delta^{(4)}\left(\sum_i p_i -\sum_f p_f\right)\, {\cal A} \ .
\ee
We study two kinematic limits: the low-energy nonrelativistic limit (this Section) and high-energy ultrarelativistic limit (next Section). In the CM frame the amplitudes are functions of the polar angle $\theta$ describing the direction of the outgoing momentum, as well as  $|\vec{p}|/m_e$, the magnitude of the  3-momentum rescaled by the electron mass $m_e$. 

In the following, we choose our initial state to be each one of the 60 stabilizer states $|\psi_\ss\>_{i}$, $i = 1,2, \cdots, 60$, for two-qubit systems. The first 36 stabilizer states are not entangled while the rest are maximally entangled. A full list of these states is given in Appendix \ref{app:ss2q}. At a particular angle $\theta$, the amplitude gives the probability of the spin configuration of the outgoing particles, represented by a state vector $|\tilde{\psi}\>$ in the computational basis: $\{|00\>,|01\>,|10\>,|11\>\}$, which is matched to $\{ |\uaua\>, |\uada\>, |\daua \>, |\dada\> \}$. We use the normalized state $|\psi\> = |\tilde{\psi}\>/\sqrt{\<\tilde{\psi}|\tilde{\psi}\>}$ to calculate the magic of the outgoing particles at this particular $\theta$, for a given initial stabilizer state.

\subsection{$e^-e^+\to \mu^-\mu^+$}
\begin{figure}
    \centering
    \includegraphics[width=0.8\linewidth]{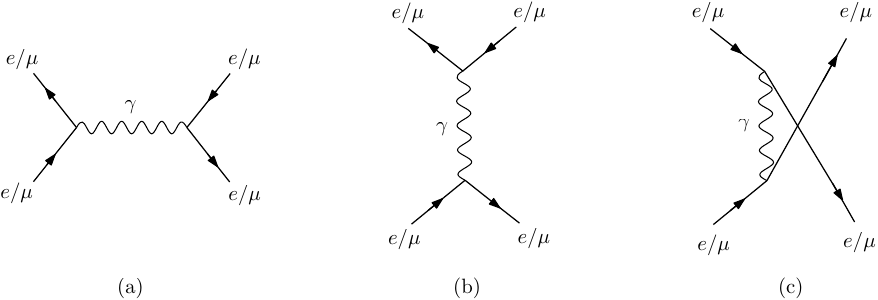}
    \caption{Representative tree-level Feynman diagrams in QED processes involving muons and electrons: (a) $s$-channel, (b) $t$-channel and (c) $u$-channel.}
    \label{fig:muon-diagram}
\end{figure}

The Feynman diagram for  $e^-e^+\to \mu^-\mu^+$ only has the $s$-channel contribution shown in Fig.~\ref{fig:muon-diagram}. The amplitude is non-vanishing only when the CM energy $\sqrt{s}$ is above the kinematic threshold $\sqrt{s}\ge 2m_\mu$. At the threshold $|\vec{p}|/m_\mu \to \sqrt{1-\lambda^2}$, where $\lambda \equiv m_e/m_\mu$, the amplitude reduces to just constant, without any $\theta$ dependence, 
\begin{align}
\label{eq:1}
\mathcal{A}_{\left|\uparrow\downarrow\right\rangle\to \left|\uparrow\downarrow\right\rangle} &= \mathcal{A}_{\left|\downarrow\uparrow\right\rangle\to \left|\downarrow\uparrow\right\rangle} = 
  - \lambda, \\
    \label{eq:2}
    \mathcal{A}_{\left|\uparrow\downarrow\right\rangle\to \left|\downarrow\uparrow\right\rangle} &= \mathcal{A}_{\left|\downarrow\uparrow\right\rangle\to \left|\uparrow\downarrow\right\rangle} =   \lambda, \\
    \mathcal{A}_{\left|\uparrow\uparrow\right\rangle\to \left|\uparrow\uparrow\right\rangle} &= \mathcal{A}_{\left|\downarrow\downarrow\right\rangle\to \left|\downarrow\downarrow\right\rangle} = - 2 \label{eq:3},
\end{align}
In the above an overall factor of $e^2$ would appear in the amplitude. However, such a factor drops out when we calculate the magic using the normalized state vector $|\psi\>$. Hence we ignore the coupling constants throughout this work. 

Interestingly, of the 60 stabilizer states, there is one state $| \psi_\ss\>_{41} = (|\uada \> + | \daua \>)/\sqrt{2}$ whose amplitude vanishes identically, at all CM energies,
\bea
\amp (| \psi_\ss\>_{41}) = 0,
\eea
with $| \psi_\ss\>_{41} = (|\uada \> + | \daua \>)/\sqrt{2}$.

\begin{table}[tbp]
    \centering
    \begin{tabular}{|c|c|c|c|}
    \hline
       {Stabilizer States} & $\ \ \Xi_2\ \ $ & $(M_2)_{\max}$ & $\lambda_{\max}$ \\
         \hline \hline
     $1,2,3,4,5,6,9,10,37,38,39,40,42,43,44,45,48,49,50$ & $\fun_1$ & ---  & ---\\
 \hline
      $7,8,11,12,46,47,59,60$   & $\flm_1$ & $\log (4/3)$ & $\sqrt{2} - 1$ \\
\hline
     \makecell{$13,14,15,16,17,18,19,20,21,22,23,24,25,26,27,28,$ \\ $ 29,30,
31,32,33,34,35,36,51,52,53,54,55,56,57,58$}    & $\flm_2$ & $\log (9/5)$ & 1 \\
     \hline
    $41$     & --- & --- & --- \\
    \hline
    \end{tabular}
    \caption{$\Xi_2$ for $e^-e^+\to \mu^-\mu^+$ at the threshold. }
    \label{tab:muple}
\end{table}

\begin{figure}[tbp]
    \centering
    \includegraphics[width=0.45\linewidth]{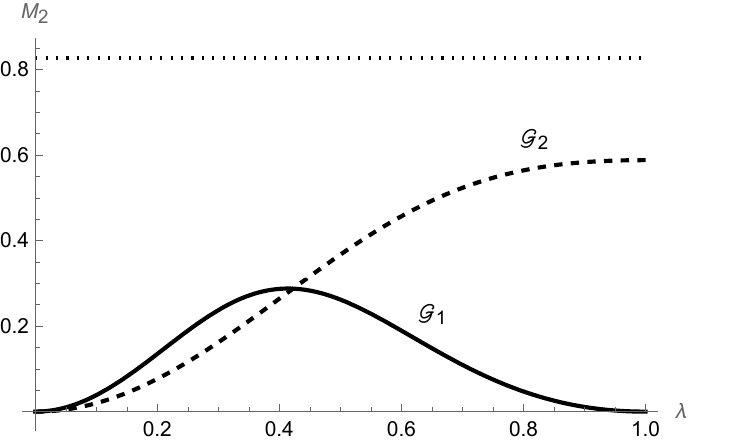}
    \caption{The final state magic as a function of $\lambda = m_e/m_\mu$ in the low energy muon production. The dotted line is $\log (16/7)$, the maximal magic for two-qubit.}
    \label{fig:g1g2}
\end{figure}

Among the remaining 59 stabilizer states, we compute $\Xi_2$ of the final states and find that it is given by only three possible values:
\bea
\fun_1 = 1, \qquad \flm_1 =  \frac{\lambda^8 + 14 \lambda^4 + 1}{(\lambda^2 + 1)^4},  \qquad \flm_2 = \frac{\lambda^8 + 28 \lambda^4 + 16}{(\lambda^2 + 2)^4} \ .
\eea
Throughout the paper we will use $\fun$ to denote functions of $\theta$ with no $\lambda$ dependence, and $\flm$ for functions with $\lambda$ dependence. Table~\ref{tab:muple} presents the final state magic resulting from each of the stabilizer initial states; for the cases with $\lambda$ dependence, we show the largest magic $(M_2)_{\max}$ and the corresponding mass ratio $\lambda_{\max}$ within $\lambda \in [0,1]$. In Fig. \ref{fig:g1g2} we plot the SRE $M_2$ as a function of $\lambda$ for the cases of $\Xi_2 = \flm_{1,2}$. If we plug in the real world values of $m_e$ and $m_\mu$, we have $\lambda \approx 0.005$, thus $\flm_{1,2} \approx 1$. Therefore, for all of the stabilizer states as the initial state, the production of magic is  minuscule:
\bea
M_2 (|\psi_\ss\>_1 ) &=& - \log \fun_1 = 0, \qquad M_2 (|\psi_\ss\>_7 ) =-\log \flm_1 \approx 9 \times 10^{-5} , \non\\
M_2 (|\psi_\ss\>_{13} ) &=& -\log \flm_2 \approx 5 \times 10^{-5}. 
\eea
On the other hand, if we vary $\lambda$, the largest magic one can achieve is through $\flm_2$, with $(M_2)_{\max} = \log (9/5) \approx 0.588$ when $\lambda = 1$.

\subsection{M{\o}ller scattering $e^-e^-\to e^-e^-$}
\label{sec:moele}

\begin{table}[tbp]
    \centering
    \begin{tabular}{|c|c|c|c|}
    \hline
        Stabilizer States &\ $\Xi_2$ \ & $(M_2)_{\max}$ & $\theta_{\max}$ \\
         \hline \hline
     $1,2,5,6,9,10,37,38,39,40,41,42,45,48,49,50$ & $\fun_1$ & 0 & Arbitrary\\
 \hline
      $3,4,7,8,11,12,43,44,46,47,59,60$    & $\fun_2$ & $\log (4/3)$& \makecell{$2 \arctan \sqrt{\sqrt{2} - 1}$,\\
     $\pi - 2 \arctan \sqrt{\sqrt{2} - 1}$}\\
     \hline
      \makecell{$13,14,15,16,17,18,19,20,21,22,23,24,25,26,27,28,$ \\ $ 
29,30,31,32,33,34,35,36,51,52,53,54,55,56,57,58$}   & $\fun_3$  & $\log (9/5)$ & $\arctan 2 \sqrt{2}$, $\pi - \arctan 2 \sqrt{2}$\\
\hline
    \end{tabular}
    \caption{$\Xi_2$ produced in the low energy limit of M{\o}ller scattering, as well as the largest magic generated in each case.}
    \label{tab:moele}
\end{table}
 
\begin{figure}
    \centering
    \includegraphics[width=0.45\linewidth]{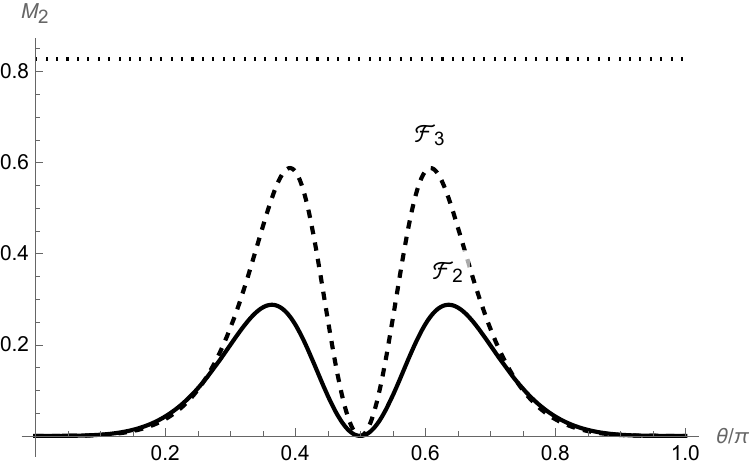}
    \caption{The distributions of the final state magic appearing in the low energy M{\o}ller $e^-e^-\to e^-e^-$ scattering. The dotted line is $\log (16/7)$, the maximal magic for two-qubit.}
    \label{fig:f2f3}
\end{figure}

The relevant Feynman diagrams are the $t$- and $u$-channel diagrams  in Fig.~\ref{fig:muon-diagram}. In the low energy regime, we  expand the amplitude in $\mu = {|\vec{p}|}/{m_e}$ and  the leading order terms scale as $1/\mu^2$, which introduces non-trivial $\theta$ dependence:
\begin{align}    \mathcal{A}_{\left|\uparrow\uparrow\right\rangle\to \left|\uparrow\uparrow\right\rangle} &= \mathcal{A}_{\left|\downarrow\downarrow\right\rangle\to \left|\downarrow\downarrow\right\rangle} =   \frac{4 \cot\theta  \csc\theta }{\mu^2}\ , 
\\
\mathcal{A}_{\left|\uparrow\downarrow\right\rangle\to \left|\uparrow\downarrow\right\rangle} &= \mathcal{A}_{\left|\downarrow\uparrow\right\rangle\to \left|\downarrow\uparrow\right\rangle} = \frac{\csc^2({\theta }/{2})}{\mu^2}\ , \\
    \mathcal{A}_{\left|\uparrow\downarrow\right\rangle\to \left|\downarrow\uparrow\right\rangle} &= \mathcal{A}_{\left|\downarrow\uparrow\right\rangle\to \left|\uparrow\downarrow\right\rangle} = - \frac{\sec^2({\theta }/{2})}{\mu^2}\ .
\end{align}
For the 60 stabilizer initial states, the corresponding final state magic  can be categorized into three groups, one has zero magic ($\Xi_2=1$) for any $\theta$ angle, and the other two 
has the following forms for $\Xi_2$:
\bea
\fun_2 = \frac{16 \left(\cos ^8\theta +14 \cos ^4\theta +1\right)}{(\cos \
2 \theta +3)^4} , \qquad \fun_3 = 
\frac{993 \cos 2 \theta +294 \cos 4 \theta +15 \cos 6 \theta +746}{4 (3 \cos 2 \theta +5)^3}\ .\ 
\eea
Notice that, similar to the coupling constant $e$, the overall factor of $\mu^2$ in the amplitudes disappears when using the normalized state vector to compute magic. The correspondence between the initial states $|\psi_\ss\>_i$ and different forms of $\Xi_2$ is given in Table \ref{tab:moele}. The distribution of $M_2$ corresponding to $\fun_{2,3}$ is plotted in Fig. \ref{fig:f2f3}, which also shows the largest magic attained in this process $(M_2)_{\max}$ and the corresponding polar angle $\theta_{\max}$. We see that the largest  magic produced in low-energy M{\o}ller scattering  is given by $\fun_3$ with $(M_2)_{\max} = \log (9/5) \approx 0.588$.

\subsection{Bhabha scattering $e^-e^+\to e^-e^+$}
The relevant Feynman diagrams are the $s$- and $t$-channels in  Fig.~\ref{fig:muon-diagram}. Similar to M{\o}ller scattering, the leading order amplitude of Bhabha scattering at low energies also scales as $1/\mu^2$. Moreover, the amplitude is proportional to identity in the spin space,\footnote{This implies the spin-entanglement is suppressed in Bhabha scattering \cite{Low:2021ufv}.}  
\begin{equation}
    \mathcal{A} = \frac{2}{\mu^2(1-\cos\theta)} + \mathcal{O}(\mu^0)
\end{equation}
Therefore, the scattering will not change the initial spin configuration, and for all stabilizer states as initial states, the final state  remains a stabilizer state and has zero magic.

\subsection{$e^-\mu^-\to e^-\mu^-$}

The relevant Feynman diagram is only the $t$-channel in Fig.~\ref{fig:muon-diagram}.
Similar to Bhabha scattering, we again find that the leading order amplitudes scale as $1/\mu^2$, and that the whole amplitude is proportional to the identity gate in the spin space,
\begin{equation}
    {\cal A} = \frac{2\lambda}{\mu^2 (-1+\cos\theta)}
    +{\cal O}(\mu^0).
\end{equation}
Therefore, no magic will be generated in $e^-\mu^-\to e^-\mu^-$ scattering at tree level.

\subsection{$\mu^-\mu^+\to e^-e^+$}

$\mu^-\mu^+\to e^-e^+$ is similar to its inverse process in that they both involve only one $s$-channel diagram at tree level, and their high energy behavior is the same. However, $\mu^-\mu^+\to e^-e^+$ does not have a kinematic threshold, and to study its low-energy behavior, we take the non-relativistic limit for the incoming muons, $|\vec{p}|/m_\mu \to0$, which leads to a different amplitude, 
\begin{align}
\mathcal{A}_{\left|\uparrow\uparrow\right\rangle\to \left|\uparrow\uparrow\right\rangle}  =  & \mathcal{A}_{\left|\downarrow\downarrow\right\rangle\to \left|\downarrow\downarrow\right\rangle}  = \frac{1}{2} (-3-\lambda +(-1+\lambda ) \cos2 \theta ) \ , \label{eq:mutoeamp1} \\
\mathcal{A}_{\left|\uparrow\uparrow\right\rangle\to \left|\downarrow\downarrow\right\rangle} =& \mathcal{A}_{\left|\downarrow\downarrow\right\rangle\to \left|\uparrow\uparrow\right\rangle} = \left(1-\lambda\right) \sin^2\theta \ ,  \label{eq:mutoeamp2}\\
\mathcal{A}_{\left|\uparrow\uparrow\right\rangle\to \left|\uparrow\downarrow\right\rangle} = &  - \mathcal{A}_{\left|\uparrow\uparrow\right\rangle\to \left|\downarrow\uparrow\right\rangle}  = \mathcal{A}_{\left|\downarrow\downarrow\right\rangle\to \left|\uparrow\downarrow\right\rangle}  = -\mathcal{A}_{\left|\downarrow\downarrow\right\rangle\to \left|\downarrow\uparrow\right\rangle} \nonumber  \\
= \mathcal{A}_{\left|\uparrow\downarrow\right\rangle\to\left|\uparrow\uparrow\right\rangle } & =   -\mathcal{A}_{\left|\downarrow\uparrow\right\rangle\to\left|\uparrow\uparrow\right\rangle }  = \mathcal{A}_{\left|\uparrow\downarrow\right\rangle\to\left|\downarrow\downarrow\right\rangle }  = -\mathcal{A}_{\left|\downarrow\uparrow\right\rangle\to\left|\downarrow\downarrow\right\rangle } =
(1-\lambda ) \cos\theta  \sin\theta \  , 
  \label{eq:mutoeamp3}   \\
\mathcal{A}_{\left|\uparrow\downarrow\right\rangle\to\left|\uparrow\downarrow\right\rangle}  = &  -\mathcal{A}_{\left|\downarrow\uparrow\right\rangle\to\left|\uparrow\downarrow\right\rangle }  = -\mathcal{A}_{\left|\uparrow\downarrow\right\rangle\to\left|\downarrow\uparrow\right\rangle}  = \mathcal{A}_{\left|\downarrow\uparrow\right\rangle\to\left|\downarrow\uparrow\right\rangle} =     \frac{1}{2} \left(-1-\lambda + (1-\lambda ) \cos2 \theta \right) \ . \label{eq:mutoeamp4} 
\end{align}
The relation between the initial states $|\psi_s\>_i$ and different forms of $\Xi_2$ is given in Table \ref{tab:mmee}.

\begin{table}[tbp]
    \centering
    \begin{tabular}{|c|c|}
    \hline
        Stabilizer States & \ $\Xi_2$\  \\
         \hline \hline
     $1,2,39,40$ & $\flm_3$  \\
 \hline
      $3,4,42,43,44$ & $\flm_{4} $  \\
     \hline
          $5,6,37,49,50$    & $\flm_{5}$ \\
     \hline
          $7,8,59,60$ & $\flm_{6}$ \\
     \hline
     $9,10,38,45,48$ & $\fun_{1}$ \\
     \hline
          $11,12$    & $\flm_7$ \\
     \hline     
     $13,14,15,16,17,18,19,20,21,22,23,24,25,26,27,28$ & $ \flm_{8} $ \\
     \hline
     $29,30,31,32,52,54,57,58$   & $\flm_{9}$  \\
\hline
     $33,34,35,36,51,53,55,56$ & $\tilde{\flm}_{9}$ \\
     \hline
     $41$ & --- \\
     \hline
     $46$ & $\flm_{10}$ \\
     \hline
     $47$ & $\tilde{\flm}_{10}$ \\
     \hline
    \end{tabular}
    \caption{$\Xi_2$ for the low energy limit of $\mu^-\mu^+ \to e^-e^+$, with $\tilde{\flm}_i (\theta) \equiv \flm_i (\pi - \theta)$.}
    \label{tab:mmee}
\end{table}

\begin{figure}[htbp]
 \centering
    \subfloat[\centering $\flm_3$ ]{{\includegraphics[width=0.3\textwidth]{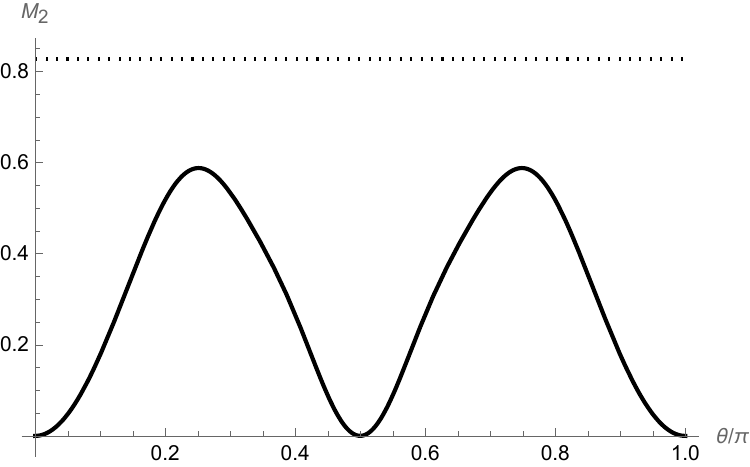} }}%
    \quad
    \subfloat[\centering $\flm_4$ ]{{\includegraphics[width=0.3\textwidth]{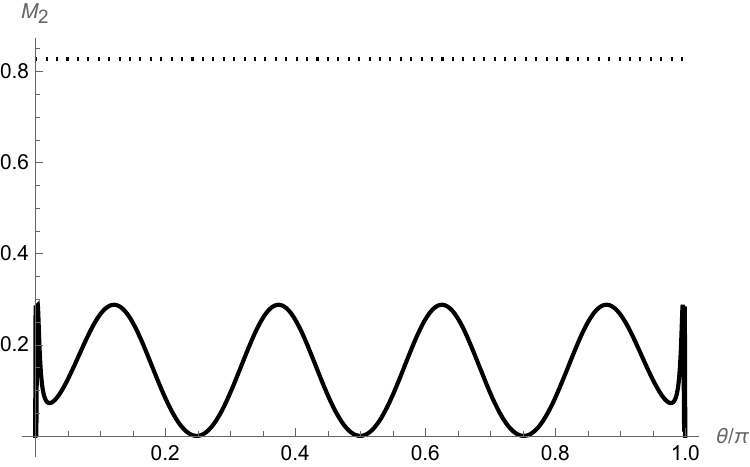} }} 
    \quad
        \subfloat[\centering $\flm_{5}$ ]{{\includegraphics[width=0.3\textwidth]{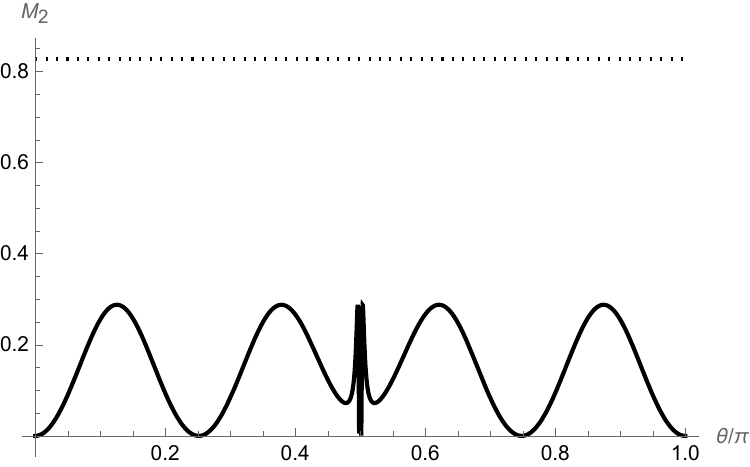} }}%
    \\
    \subfloat[\centering $\flm_{6}$ ]{{\includegraphics[width=0.3\textwidth]{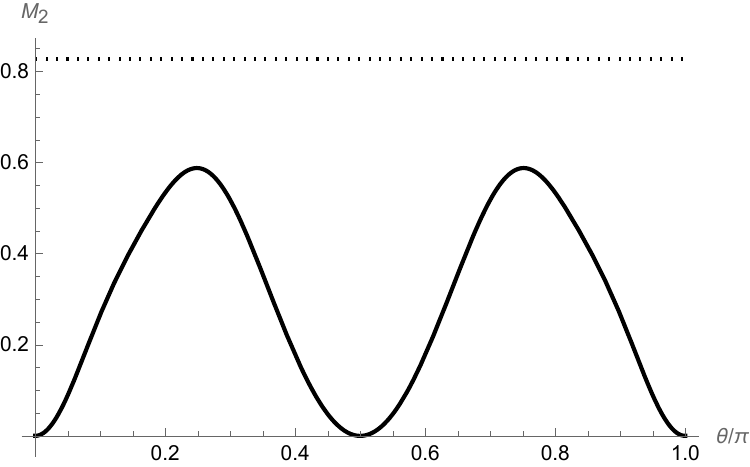} }} 
    \quad
            \subfloat[\centering $\flm_{7}$ ]{{\includegraphics[width=0.3\textwidth]{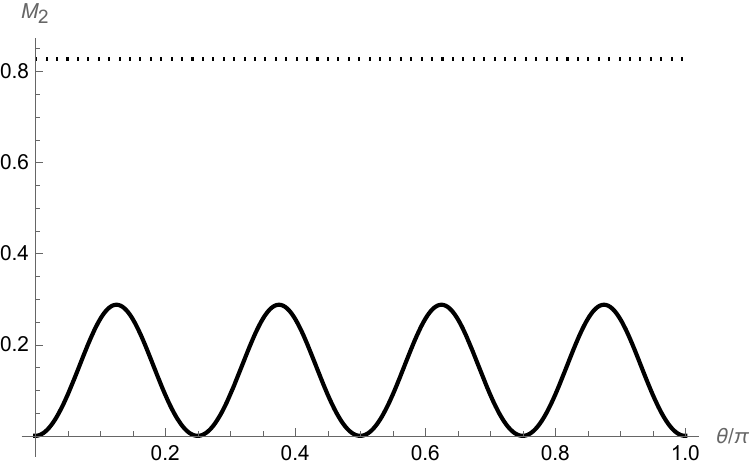} }}%
    \quad
    \subfloat[\centering $\flm_{8}$ ]{{\includegraphics[width=0.3\textwidth]{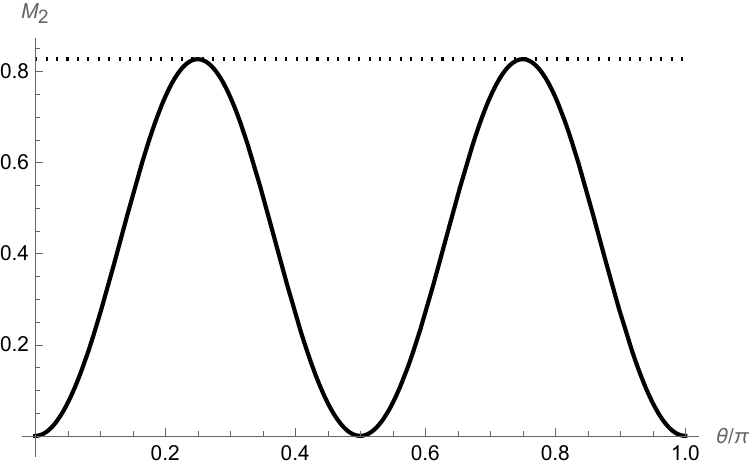} }} \\
            \subfloat[\centering $\flm_{9}$ ]{{\includegraphics[width=0.3\textwidth]{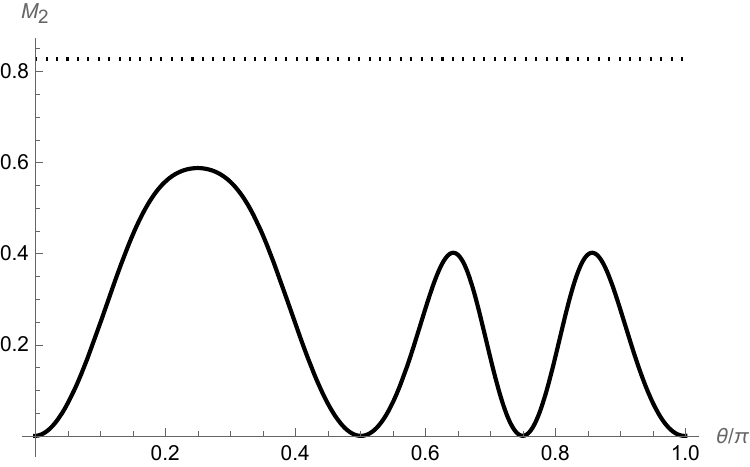} }}%
    \quad
    \subfloat[\centering $\flm_{10}$ ]{{\includegraphics[width=0.3\textwidth]{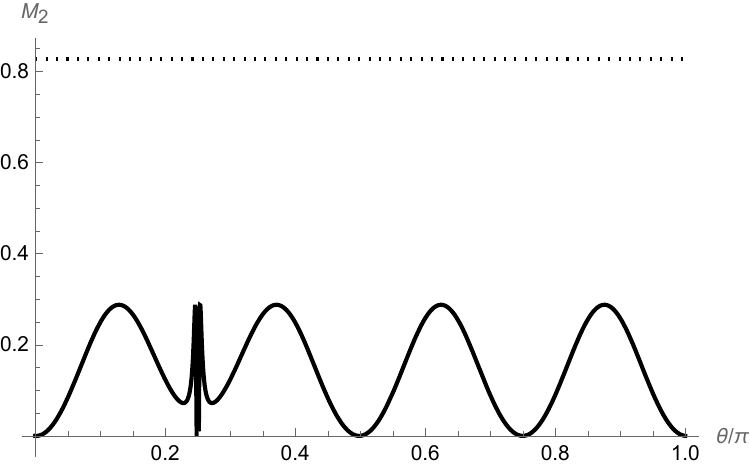} }} 
    \caption{Distributions of $M_2$ in  low energy $\mu^-\mu^+ \to e^-e^+$ with $\lambda=m_e/m_\mu$ set to the real world value of $\sim 0.005$. The dotted line is $\log (16/7)$, the maximal magic for two-qubit. \label{fig:ff8}}
\end{figure}

\begin{figure}[htbp]
    \centering
    \includegraphics[width=0.45\linewidth]{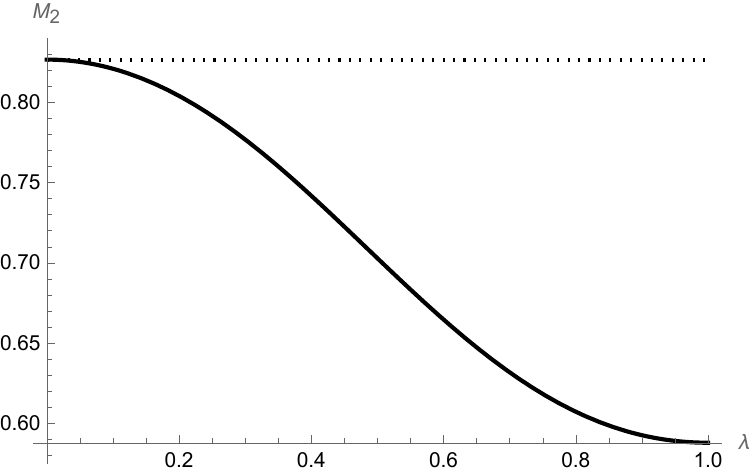}
    \caption{The maximal magic achieved from initial states $|\psi_\ss\>_{i}$, $i=13,14, \cdots ,28$, in the low energy scattering of $\mu^-\mu^+ \to e^-e^+$. These maximal values are attained at $\theta = \pi/4$ or $3\pi/4$.}
    \label{fig:g8m}
\end{figure}

For general values of $\lambda =m_e/m_\mu \in [0,1]$, we have $\flm_i (\theta) = \flm_i (\pi - \theta)$ for $i = 3,4,5,6,7,8$, $\flm_5 (\theta) = \flm_4 (\pi/2 - \theta)$, $\flm_6 (\theta) = \flm_3 (\pi/2 - \theta)$ and $\flm_{10} (\theta) = \flm_4 (\theta - \pi/4)$. The functions $\flm_i$, $i = 3,4,7,8,9 $ are presented in machine readable format in the supplementary material of this publication \cite{supp},  and the  distributions are shown in Fig.~\ref{fig:ff8}. We notice that $\flm_8$ results in significantly higher largest magic than any other distributions. More specifically, 
\bea
\left[M_2 (|\psi_\ss\>_{13} )\right]_{\max} &=& -\log \min (\flm_8 ) = -\log \frac{\lambda ^8+19 \lambda ^4+18 \lambda ^2+7}{\left(\lambda ^2+2\right)^4},
\eea
which occurs at $\theta = \pi/4$ or $3\pi/4$. We plot the above as a function of $\lambda$ in Fig. \ref{fig:g8m}. For $\lambda = 0$, the above gives $\log (16/7)$ which is the maximal magic that can be realized for a 2-qubit system; the minimal value of the above is $\log (9/5)$, attained at $\lambda = 1$. For the mass ratio $\lambda \approx 0.005$ in the real world, $\left[M_2 (|\psi_\ss\>_{13} )\right]_{\max}$ is very close to $\log (16/7)$.

\section{Magic in QED at High Energies}
\label{sect:4}

\subsection{$e^-e^+\to \mu^-\mu^+$}
The ultrarelativistic limit is given by $\mu = |\vec{p}|/m_e \to \infty$ limit. Thus we will expand the amplitude in power series of $1/\mu$ and present the analytic expressions up to ${\cal O}(1/\mu)$. The reason being that, for some stabilizer initial states, the leading order amplitude is ${\cal O}(1)$, while for some others it is ${\cal O}(1/\mu)$:
\begin{align}
\mathcal{A}_{\left|\uparrow\uparrow\right\rangle\to \left|\uparrow\downarrow\right\rangle} & =  - \mathcal{A}_{\left|\uparrow\uparrow\right\rangle\to \left|\downarrow\uparrow\right\rangle}  = \mathcal{A}_{\left|\downarrow\downarrow\right\rangle\to \left|\uparrow\downarrow\right\rangle}  = -\mathcal{A}_{\left|\downarrow\downarrow\right\rangle\to \left|\downarrow\uparrow\right\rangle} = \frac{1}{2} \left(1-\frac{1}{\mu}\right) \sin2\theta \ , \\
    \mathcal{A}_{\left|\uparrow\uparrow\right\rangle\to \left|\downarrow\downarrow\right\rangle} &= \mathcal{A}_{\left|\downarrow\downarrow\right\rangle\to \left|\uparrow\uparrow\right\rangle} = \frac{1}{2}\left(1-\frac{1}{\mu}\right) (1-\cos2\theta ) \ , \\
    \mathcal{A}_{\left|\uparrow\uparrow\right\rangle\to \left|\uparrow\uparrow\right\rangle} &= \mathcal{A}_{\left|\downarrow\downarrow\right\rangle\to \left|\downarrow\downarrow\right\rangle} =  -\frac{1}{2} (3+\cos2\theta ) +\frac{1}{2\mu}(-1+\cos2\theta ) \ ,  \\
    \mathcal{A}_{\left|\uparrow\downarrow\right\rangle\to\left|\uparrow\uparrow\right\rangle } & =   -\mathcal{A}_{\left|\downarrow\uparrow\right\rangle\to\left|\uparrow\uparrow\right\rangle }  = \mathcal{A}_{\left|\uparrow\downarrow\right\rangle\to\left|\downarrow\downarrow\right\rangle }  = -\mathcal{A}_{\left|\downarrow\uparrow\right\rangle\to\left|\downarrow\downarrow\right\rangle } = \frac{\lambda}{2\mu} \sin2\theta\ ,\label{eq:mupahe4} \\
    \mathcal{A}_{\left|\uparrow\downarrow\right\rangle\to\left|\uparrow\downarrow\right\rangle} & =   -\mathcal{A}_{\left|\downarrow\uparrow\right\rangle\to\left|\uparrow\downarrow\right\rangle }  = -\mathcal{A}_{\left|\uparrow\downarrow\right\rangle\to\left|\downarrow\uparrow\right\rangle}  = \mathcal{A}_{\left|\downarrow\uparrow\right\rangle\to\left|\downarrow\uparrow\right\rangle} = - \frac{\lambda}{2\mu} (1-\cos2\theta ) .\label{eq:mupahe5} 
\end{align}
For $|\psi_{\text{s}}\>_i$ with $i = 3,4,41,42,43,44$, the amplitude   vanishes at $\mathcal{O} (\mu^0)$ and we need to include the ${\cal O}(1/\mu)$ contribution. For example, if the initial state is $|\psi_\text{s} \>_3 = |\uada\>$, the final state is given by Eqs. (\ref{eq:mupahe4}) and (\ref{eq:mupahe5}), which start at $\mathcal{O} (1/\mu)$. Again the overall factor of $1/\mu$ cancels out when using a normalized final state to calculate the magic.

\begin{figure}[tbp]
    \centering
    \includegraphics[width=0.45\linewidth]{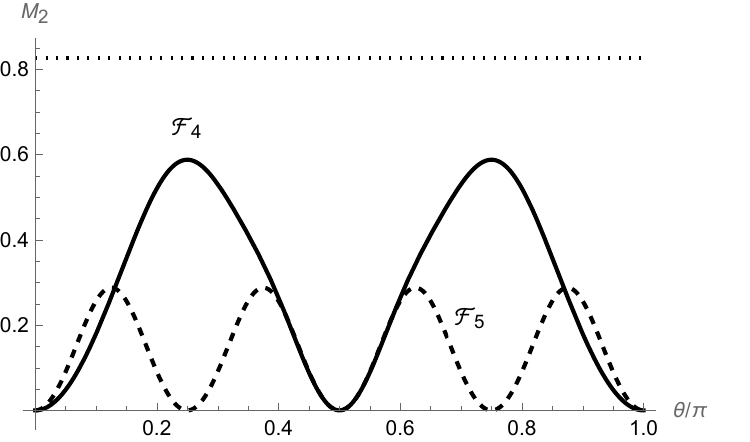}
    \caption{The distributions of the final state magic appearing in the high energy scattering of $e^-e^+\to \mu^-\mu^+$. The dotted line is $\log 16/7$, the maximal magic for two-qubit.}
    \label{fig:f4f5}
\end{figure}

The functions entering $\Xi_2$ in this case are,
\bea
\fun_4&=& \frac{1}{128 (\cos 2 \theta +3)^4} \Big[13336 \cos 2 \theta +5796 \cos 4 \theta +1960 \cos 6 \theta +532 \cos 8 \theta +56 \cos 10 \theta\non\\
&&+28 \cos 12 \theta +8 \cos 14 \theta +\cos 16 \theta +11051 \Big],\\
\fun_5 &=& \frac{1}{8} (\cos 8 \theta +7) ,
\eea
and Table \ref{tab:muphe} gives the corresponding initial states. The magic produced from  $\fun_{4,5}$ is shown in Fig.~\ref{fig:f4f5}. We note that $ \flm_3 = \fun_4$ and $ \flm_{4,5,7,10} = \fun_5$ when $\lambda = 0$. Similar to the low energy M{\o}ller scattering discussed in Section \ref{sec:moele}, the  maximal magic generated is also $\log (9/5)$.

\begin{table}[tbp]
    \centering
    \begin{tabular}{|c|c|c|c|}
    \hline
        Stabilizer States &\ $\Xi_2$\ & $(M_2)_{\max}$ & $\theta_{\max}$ \\
         \hline \hline
     \makecell{$1,2,13,14,15,16,17,18,19,20,21,22,23,24,25,26,27,28,$ \\ $ 29,30,31,32,33,34,35,36,39,40,51,52,53,54,55,56,57,58$ } & $\fun_4$ & $\log (9/5)$  & $\pi/4$, $3\pi/4$ \\
 \hline
     $3,4,5,6,11,12,37,42,43,44,46,47,49,50$   & $\fun_5$  & $\log (4/3) $& \makecell{$\pi/8$, $3\pi/8$, \\
     $5\pi/8$, $7\pi/8$}\\
\hline
     $7,8,9,10,38,45,48,59,60$    & $\fun_1$ & 0& Arbitrary \\
     \hline
     $41$ & --- & --- & ---\\
     \hline
    \end{tabular}
    \caption{$\Xi_2$ in the high energy limit of $e^-e^+\to \mu^-\mu^+$.}
    \label{tab:muphe}
\end{table}

As we mentioned earlier, $\mu^- \mu^+ \to e^- e^+$ has the same high energy behavior as  $e^- e^+  \to \mu^- \mu^+$, and thus will give the same distributions of magic.

\subsection{M{\o}ller scattering $e^-e^-\to e^-e^-$}

\begin{table}[tbp]
    \centering
    \begin{tabular}{|c|c|c|c|}
    \hline
        Stabilizer States &\ $\Xi_2$\ & $(M_2)_{\max}$ & $\theta_{\max}$ \\
         \hline \hline
     $1,2,39,40$ & $\fun_6$ & $0.576\cdots$  & $(\pi/2) \pm 0.783 \cdots$ \\
 \hline
      $3,4,43,44$ & $\fun_{7} $ & $\log (16/9)$ & $\pi/4$, $3\pi/4$ \\
     \hline
          $5,6,49,50$    & $\fun_{8}$ & $\log (9/5) $ &  $\pi/4$, $3\pi/4$, $(\pi/2) \pm \text{arccot} \sqrt{2}$ \\
     \hline
          $7,8,59,60$ & $\fun_{9}$ & $0.586\cdots$ & $(\pi/2) \pm 0.781 \cdots$\\
     \hline
     
     $9,10$ & $\fun_{10}$ & $0.268\cdots$ & $(\pi/2) \pm 0.186 \cdots$\\
     \hline
          $11,12,46,47$    & $\fun_2$ & $\log (4/3)$& \makecell{$2 \arctan \sqrt{\sqrt{2} - 1}$,\\
     $\pi - 2 \arctan \sqrt{\sqrt{2} - 1}$}\\
     \hline
     
     $13,14,15,16,17,18,19,20,21,22,23,24,25,26,27,28$ & $ \fun_{11} $ & $0.458 \cdots$ & $(\pi/2) \pm 0.444 \cdots$\\
     \hline
     $29,31,34,36,55,56,57,58$   & $\fun_{12}$  & $0.539\cdots $& $0.649 \cdots$\\
\hline
     $30,32,33,35,51,52,53,54$ & $\tilde{\fun}_{12}$ & $0.539\cdots $& $\pi - 0.649 \cdots$\\
     \hline
     $37,42$ & $\fun_1$ & $0$ & Arbitrary\\
     \hline
          $38,41$   & $\fun_5$  & $\log (4/3) $& $\pi/8$, $3\pi/8$, 
     $5\pi/8$, $7\pi/8$ 
     \\
\hline
     $45$ & $\fun_{13}$ & $\log (4/3)$ & \makecell{$0.440\cdots$, $1.49\cdots$,\\
     $2.16\cdots$, $2.78\cdots$}\\
     \hline

     $48$ & $\tilde{\fun}_{13}$ & $\log (4/3)$ & \makecell{$\pi-0.440\cdots$, $\pi-1.49\cdots$,\\
     $\pi-2.16\cdots$, $\pi-2.78\cdots$}\\
     \hline
    \end{tabular}
    \caption{$\Xi_2$ for the high energy limit of M{\o}ller scattering. Notice $\tilde{\fun}_i (\theta) \equiv \fun_i (\pi - \theta)$.}
    \label{tab:moehe}
\end{table}

\begin{figure}[tbp]
 \centering
    \subfloat[\centering $\fun_6$ ]{{\includegraphics[width=0.3\textwidth]{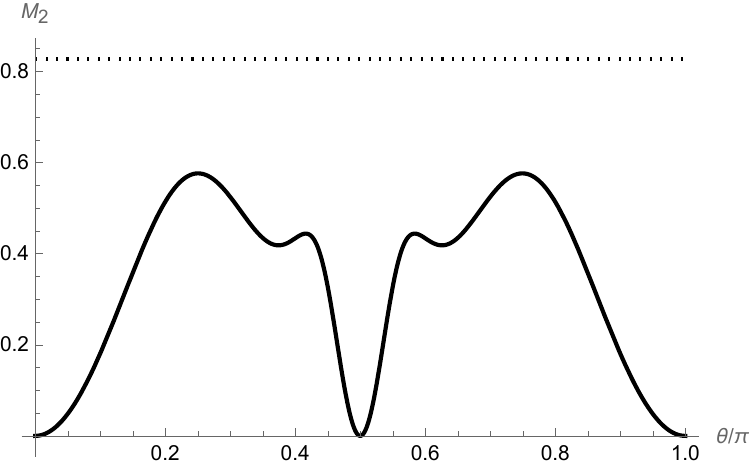} }}%
    \quad
    \subfloat[\centering $\fun_7$ ]{{\includegraphics[width=0.3\textwidth]{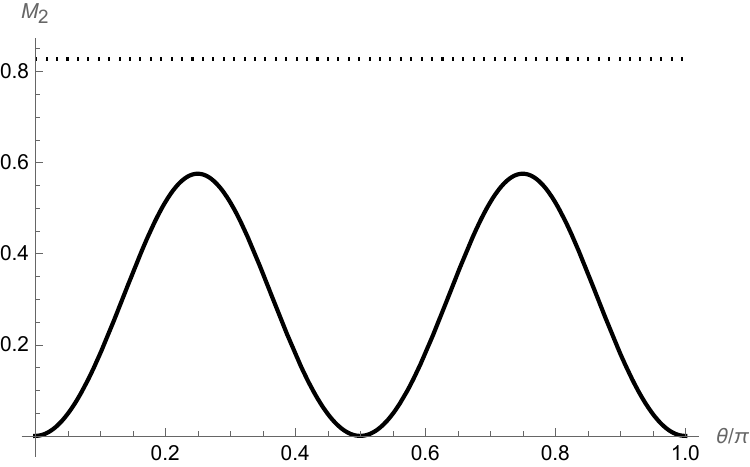} }} \quad
        \subfloat[\centering $\fun_{8}$ ]{{\includegraphics[width=0.3\textwidth]{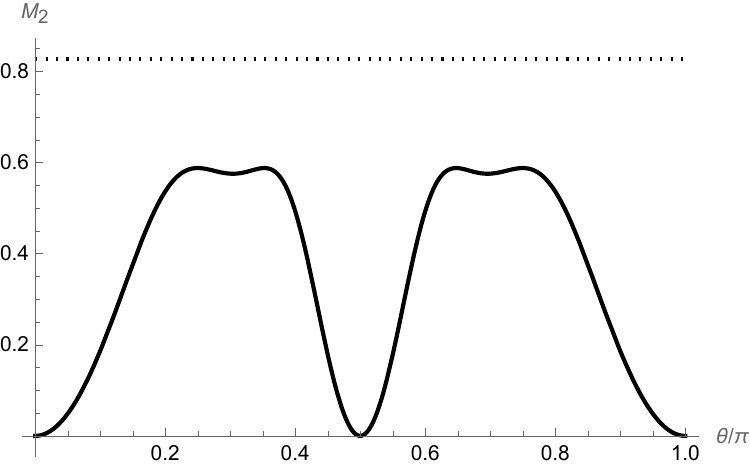} }} \\
    \subfloat[\centering $\fun_{9}$ ]{{\includegraphics[width=0.3\textwidth]{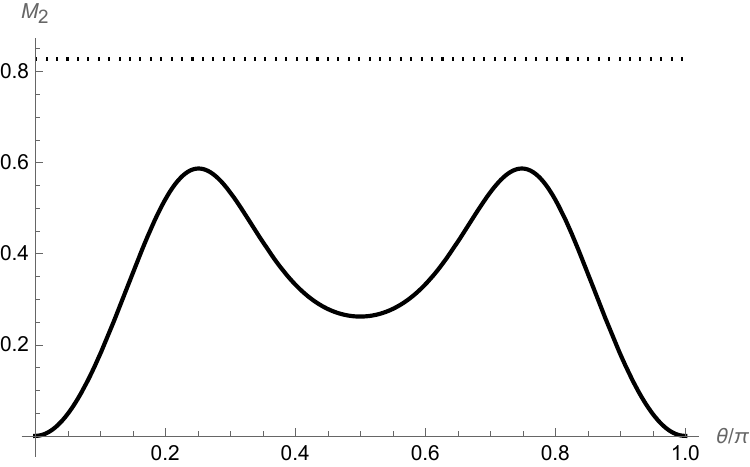} }} 
    \quad
            \subfloat[\centering $\fun_{10}$ ]{{\includegraphics[width=0.3\textwidth]{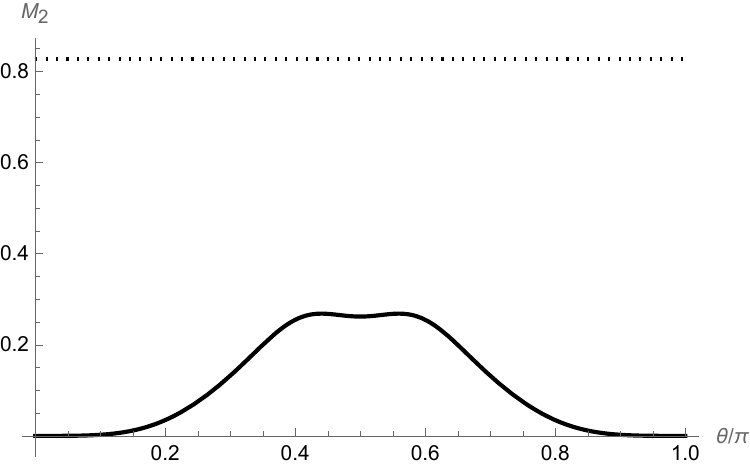} }}%
    \quad
    \subfloat[\centering $\fun_{11}$ ]{{\includegraphics[width=0.3\textwidth]{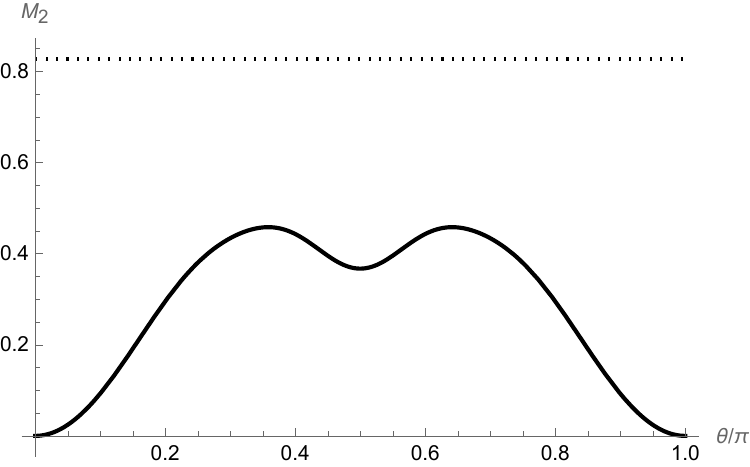} }} 
    \\
            \subfloat[\centering $\fun_{12}$ ]{{\includegraphics[width=0.3\textwidth]{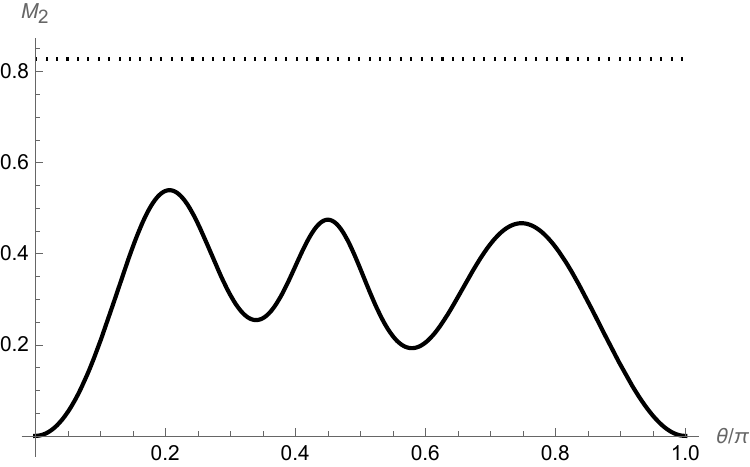} }}%
    \quad
    \subfloat[\centering $\fun_{13}$ ]{{\includegraphics[width=0.3\textwidth]{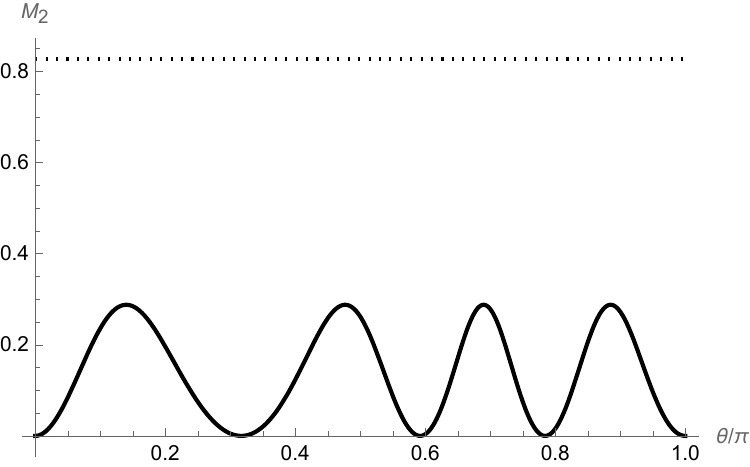} }} 
    \caption{Additional distributions of $M_2$ appearing in high energy M{\o}ller scattering. The dotted line is $\log (16/7)$, the maximal magic for two-qubit. \label{fig:f8}}
\end{figure}

The amplitude for M{\o}ller scattering in the spin space is more complicated. We show the full $4\times4$ matrix in the computational basis,
\be
  \mathcal{A} = 
\begin{pmatrix}
 (7+\cos2 \theta ) \frac{\cos\theta}{\sin^2\theta}   & (3+\cos2 \theta ) \csc\theta  & (3+\cos2 \theta ) \csc\theta
 & 2 \cos\theta  \\
 -4 \csc\theta  & 2 \csc^2 ({\theta }/{2}) & -2 \sec^2 ({\theta}/{2}) & 4 \csc\theta  \\
 -4 \csc\theta  & -2 \sec^2 \frac{\theta }{2} & 2 \csc^2 \frac{\theta }{2} & 4 \csc\theta  \\
 2 \cos\theta  & -(3+\cos2 \theta ) \csc\theta & -(3+\cos2 \theta ) \csc\theta  & (7+\cos2 \theta
) \frac{\cos\theta}{\sin^2\theta}   
\end{pmatrix}\ .
\ee
The corresponding $\Xi_2$ for each  stabilizer initial state is given in Table \ref{tab:moehe} and exhibits a much richer structure. The generating functions $\fun_j$, $j = 6, 7, \cdots $ are given in machine readable format in the supplementary material of this publication \cite{supp}. Distributions of $M_2$ that are different from previous sections are given in Fig.~\ref{fig:f8}. The 60 stabilizer states result in a total of 13 different distributions in final state magic. Curiously the  largest magic generated in the high energy limit remains $\log (9/5)$,  the same as in the low energy M{\o}ller scattering.

Also note that the magic distributions generated from $\fun_{12,13}$ are not symmetric in $\theta \to \pi - \theta$, as one would naively expect of a system with identical particles. This is because the operation of interchanging two identical particles involves not just swapping the three-momentum but also the spin states. Moreover, since we define the $z$-axis to be in the direction of motion of particle 1, swapping the three-momentum also  flips the direction of $z$-axis and induces a spin flip $|0\> \leftrightarrow |1\>$. We have verified in our computation that, when all these operations are taken into account, the amplitudes are invariant up to a minus sign, as required by the Fermi-Dirac statistics.

\subsection{Bhabha scattering $e^-e^+\to e^-e^+$}
The amplitude for high energy Bhabha scattering, shown in the computational basis, is
\begin{equation}
\mathcal{A}
= 
    \begin{pmatrix}
  \frac{15 \cos\theta +\cos3 \theta }{4-4 \cos\theta } & -\frac{1}{2} (3+\cos2 \theta ) \cot\frac{\theta
}{2} & \frac{1}{2} (3+\cos2 \theta ) \cot\frac{\theta }{2} & -2 \cos^2\frac{\theta }{2} \cos\theta
 \\
 2 \cot\frac{\theta }{2} & 2 \cot^2\frac{\theta }{2} & 2 & 2 \cot\frac{\theta }{2} \\
 -2 \cot\frac{\theta }{2} & 2 & 2 \cot^2\frac{\theta }{2} & -2 \cot\frac{\theta }{2} \\
 -2 \cos^2\frac{\theta }{2} \cos\theta  & -\frac{1}{2} (3+\cos2 \theta ) \cot\frac{\theta }{2}
& \frac{1}{2} (3+\cos2 \theta ) \cot\frac{\theta }{2} & \frac{15 \cos\theta +\cos3 \theta }{4-4 \cos\theta
}
    \end{pmatrix}\ .
\end{equation}
The final state magic is summarized in Table \ref{tab:bhahe}, which involves the same 13 distributions as the high energy M{\o}ller scattering, given in Table \ref{tab:moehe}. The largest magic is thus still $\log (9/5)$.

\begin{table}[tbp]
    \centering
    \begin{tabular}{|c|c|}
    \hline
        Stabilizer States & \ $\Xi_2$\  \\
         \hline \hline
     $1,2,39,40$ & $\fun_6$  \\
 \hline
      $3,4,43,44$ & $\fun_{7} $  \\
     \hline
          $7,8,59,60$    & $\fun_{8}$ \\
     \hline
          $5,6,49,50$ & $\fun_{9}$ \\
     \hline
     $11,12$ & $\fun_{10}$ \\
     \hline
          $9, 10, 45, 48$    & $\fun_2$ \\
     \hline     
     $13,14,15,16,17,18,19,20,21,22,23,24,25,26,27,28$ & $ \fun_{11} $ \\
     \hline
     $29,30,31,32,52,54,57,58$   & $\fun_{12}$  \\
\hline
     $33,34,35,36,51,53,55,56$ & $\tilde{\fun}_{12}$ \\
     \hline
     $38,41$ & $\fun_1$ \\
     \hline
          $37,42$   & $\fun_5$  \\
\hline
     $46$ & $\fun_{13}$ \\
     \hline
     $47$ & $\tilde{\fun}_{13}$ \\
     \hline
    \end{tabular}
    \caption{$\Xi_2$ for the high energy limit of Bhabha scattering, which involves the same 13 distributions as the high energy M{\o}ller scattering.}
    \label{tab:bhahe}
\end{table}

\subsection{$e^-\mu^-\to e^-\mu^-$}
\begin{figure}[htbp]
 \centering
    \subfloat[\centering $\fun_{14}$ ]{{\includegraphics[width=0.3\textwidth]{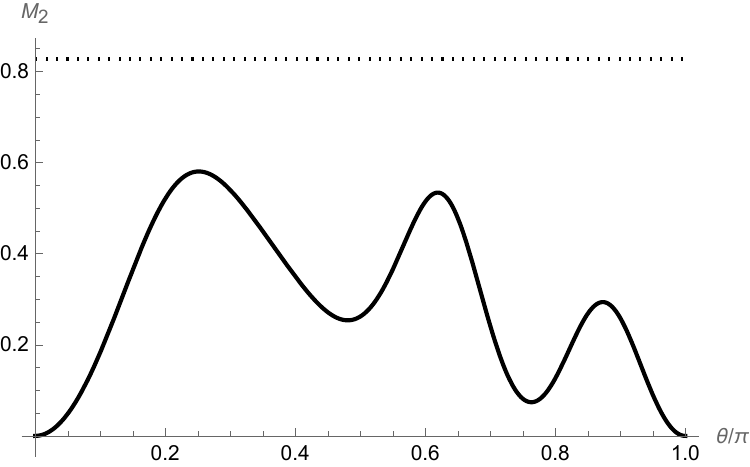} }}%
    \quad
    \subfloat[\centering $\fun_{5}$ ]{{\includegraphics[width=0.3\textwidth]{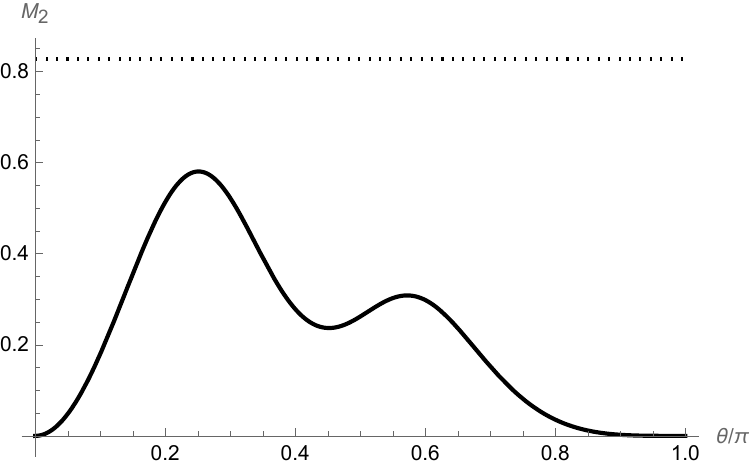} }}
    \quad
    \subfloat[\centering $\fun_{16}$ ]{{\includegraphics[width=0.3\textwidth]{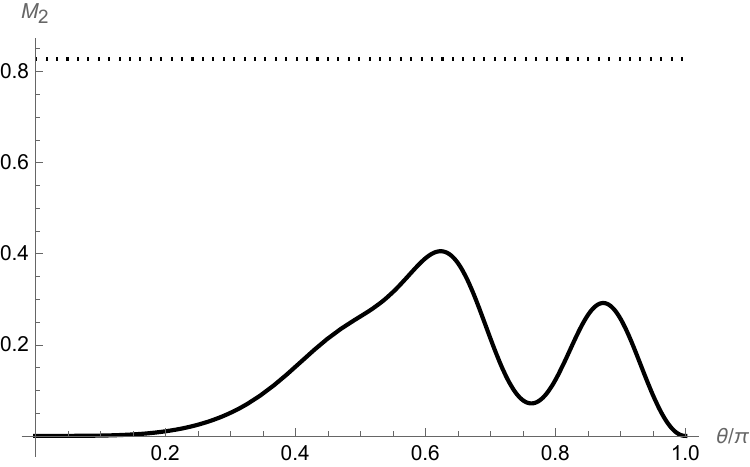} }} \\
    \subfloat[\centering $\fun_{17}$ ]{{\includegraphics[width=0.3\textwidth]{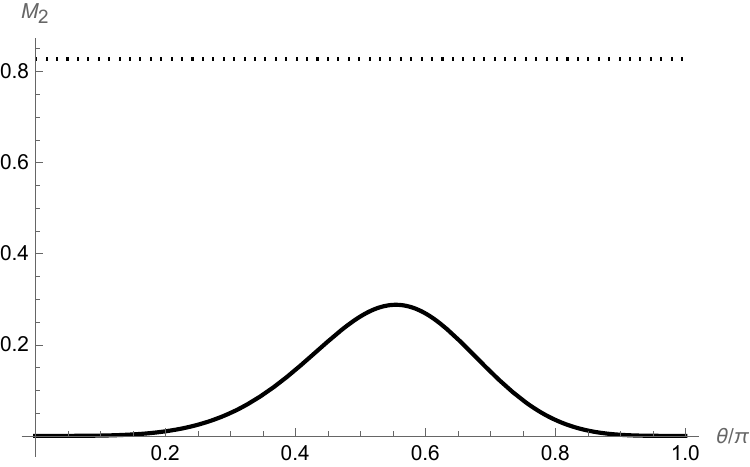} }} 
    \quad
            \subfloat[\centering $\fun_{18}$ ]{{\includegraphics[width=0.3\textwidth]{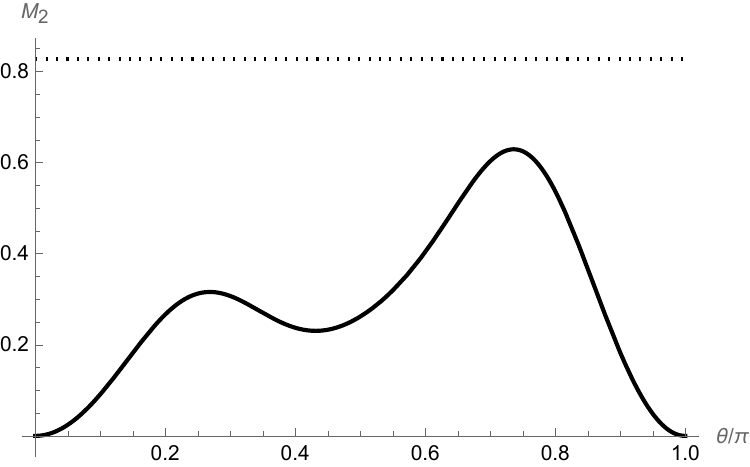} }}%
    \quad
    \subfloat[\centering $\fun_{19}$ ]{{\includegraphics[width=0.3\textwidth]{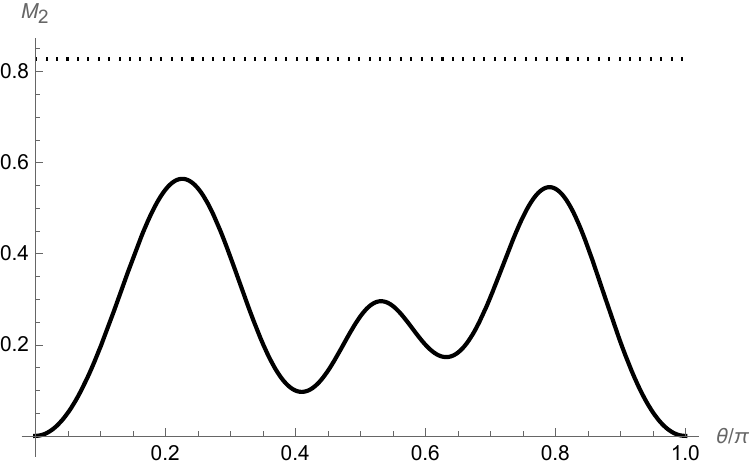} }} \\
            \subfloat[\centering $\fun_{20}$ ]{{\includegraphics[width=0.3\textwidth]{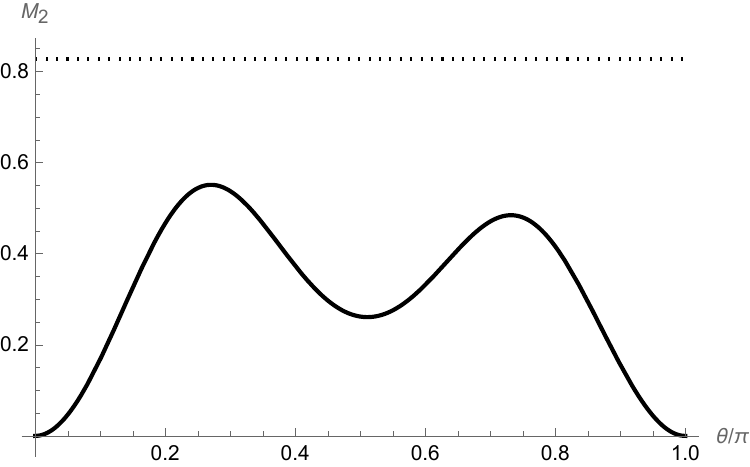} }}%
    \quad
    \subfloat[\centering $\fun_{21}$ ]{{\includegraphics[width=0.3\textwidth]{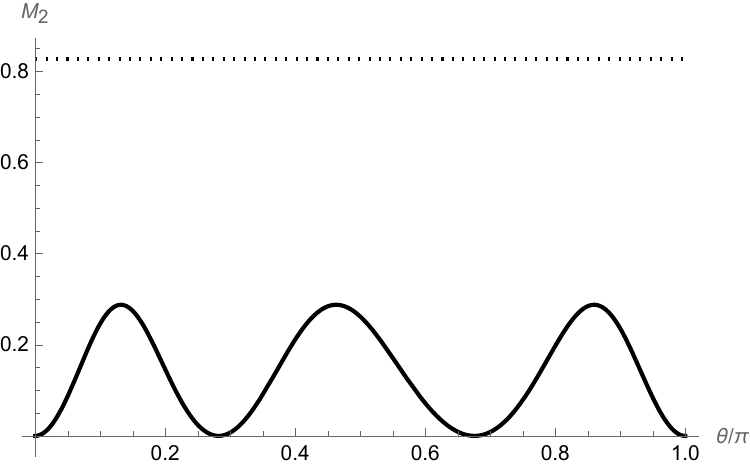} }}%
%
    \subfloat[\centering $\fun_{22}$ ]{{\includegraphics[width=0.3\textwidth]{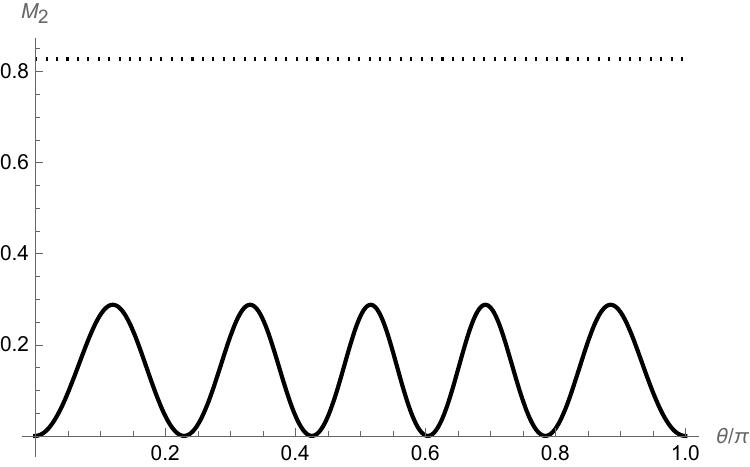} }} 
    \caption{Additional distributions of $M_2$ appearing in high energy elastic scattering of $e^-\mu^-$. The dotted line is $\log (16/7)$, the maximal magic for two-qubit. \label{fig:f16}}
\end{figure}

The high energy limit of $e^- \mu^-$ scattering has the amplitude 
\begin{equation}
\mathcal{A}
= 
    \begin{pmatrix}
3+\frac{4}{-1+\cos\theta }+\cos\theta  & -2 \cot\frac{\theta }{2}+\sin\theta  & -2 \cot\frac{\theta
}{2}+\sin\theta  & -1-\cos\theta  \\
 2 \cot\frac{\theta }{2} & -2 \cot^2\frac{\theta }{2} & 2 & -2 \cot\frac{\theta }{2} \\
 2 \cot\frac{\theta }{2} & 2 & -2 \cot^2\frac{\theta }{2} & -2 \cot\frac{\theta }{2} \\
 -1-\cos\theta  & 2 \cot\frac{\theta }{2}-\sin\theta  & 2 \cot\frac{\theta }{2}-\sin\theta
 & 3+\frac{4}{-1+\cos\theta }+\cos\theta  
    \end{pmatrix},
\end{equation}
which leads to the 12 distributions of $\Xi_2$ as given by Table \ref{tab:emuhe}.  In Fig. \ref{fig:f16} we show the additional angular distributions not shown previously.

\begin{table}[tbp]
    \centering
    \begin{tabular}{|c|c|c|c|}
    \hline
        Stabilizer States & $\Xi_2$ & $(M_2)_{\max}$ & $\theta_{\max}$ \\
         \hline \hline
     $1,2,3,4,39,40,43,44$ & $\fun_7$ & $\log (16/9)$ & $\pi/4$, $3\pi/4$ \\
 \hline
      $5,6,49,50$ & $\fun_{14} $ & $0.580\cdots$ & $0.790\cdots$ \\
     \hline
          $7,8,59,60$    & $\fun_{15}$ & $0.580\cdots$ &  $0.789\cdots$ \\
     \hline
     $9,10$ & $\fun_{16}$ & $0.405\cdots$ & $ 1.95 \cdots$\\
     \hline
          $11,12,46,47$    & $\fun_{17}$ & $\log (4/3)$& $2 \arctan 2^{1/4}$\\
     \hline
     $13,14,15,16,17,18,19,20,21,22,23,24,25,26,27,28$ & $ \fun_{18} $ & $0.628 \cdots$ & $2.31 \cdots$\\
     \hline
     $29,31,34,36,55,56,57,58$   & $\fun_{19}$  & $0.569\cdots $& $0.710 \cdots$\\
\hline
     $30,32,33,35,51,52,53,54$ & $\fun_{20}$ & $0.550\cdots $& $0.849 \cdots$\\
     \hline
     $37,42$ & $\fun_1$ & $0$ & Arbitrary\\
     \hline
          $38,41$   & $\fun_5$  & $\log (4/3) $& \makecell{$\pi/8$, $3\pi/8$, \\
     $5\pi/8$, $7\pi/8$}\\
\hline
     $45$ & $\fun_{21}$ & $\log (4/3)$ & $0.414\cdots$, $1.45\cdots$,     $2.70\cdots$\\
     \hline
     $48$ & $\fun_{22}$ & $\log (4/3)$ & \makecell{$0.375\cdots$, $1.03\cdots$, $1.62\cdots$\\
     $2.17\cdots$, $2.78\cdots$}\\
     \hline
    \end{tabular}
    \caption{$\Xi_2$ for the high energy limit of elastic $e^- \mu^-$ scattering.}
    \label{tab:emuhe}
\end{table}

\section{Discussions}

In this work we initiated a study on the ability of QED to generate quantum magic from stabilizer initial states through 2-to-2 scattering processes involving electrons and muons. We considered all 60 stabilizer states for two-qubit and found that the angular distributions for the magic of the final states are governed by only a few patterns. In most cases, the largest magic generated is significantly smaller than the maximal magic for two-qubit. Several processes do not even produce magic states at all. The only process that generates the maximal possible magic for two-qubit states is the low-energy scattering of $\mu^-\mu^+\to e^-e^+$, in the limit $\lambda=m_e/m_\mu\to 0$, which is well approximated in the real world. Our findings suggest QED is not an efficient mechanism to generate quantum advantages in quantum computation. This is a somewhat surprising result, especially given QED is highly capable of producing maximal entanglement through scattering processes.

Since the stabilizer R\`enyi entropy involves calculating the expectation values of Pauli matrices, the resulting magic is not rotationally invariant and depends on the choice of reference frame. This is easy to understand from the perspective of quantum computing, as rotating the quantum state requires additional resources, by applying quantum logic gates on the state. In order to compare the initial and final state magic on equal footing, we choose the CM frame and project the spins on the $z$-axis for both the incoming and outgoing fermions, which is different from the helicity basis projecting the spins to the direction of three-momentum. Moreover, the $x$-$z$ plane of our reference frame is defined by the outgoing particles. As such, the amplitudes and magic computed in this frame contains no dependence on the azimuthal angle $\phi$. Although our choice of reference frame can be constructed on an event-by-event basis, it is also different for every scattering event. One may wish to consider a CM frame in which the $x$- and $y$-axis are fixed and do not change on an event-by-event basis. In this case the amplitudes and magic will depend on both $\phi$ and the polar angle $\theta$. Such a more general choice will be studied in the future.

Last but no least, it is worth recalling that QED is one of the fundamental forces in nature and governs the interaction of light and matter. Our results bring into sharp focus the following question: Is any fundamental interaction in nature capable of generating maximal magic efficiently? We plan to return to this question in a future work.

\section*{Acknowledgements}
This work is supported in part by the U.S.
Department of Energy, Office of High Energy Physics,
under contract DE-AC02-06CH11357 at Argonne, and by the U.S. Department of Energy, Office of Nuclear
Physics, under grant DE-SC0023522 at Northwestern.

\appendix

\section{Stabilizer states for  two-qubit systems}
\label{app:ss2q}

We list in Table \ref{tab:stabs} the 60 stabilizer states for the 2-qubit system, in the order that we use throughout the paper. They are ordered so that the first 36 of them are products of single qubit stabilizer states, without any entanglement between the two qubits; while the rest of the states realize maximum entanglement between the two qubits.

We have chosen the spin of a particle, projected in the $z$ direction, as our single qubits, i.e. $|\uparrow\> \equiv | +z \>$ and $|\downarrow \> = | -z\>$, with $|\pm z\>$, $| \pm x\>$ and $| \pm y\>$ being spin eigenstates in the $z$, $x$ and $y$ directions. The 6 single qubit stabilizer states are exactly $|\pm z\>$, $| \pm x\>$ and $| \pm y\>$. We thus organize the first 36 of the 2-qubit stabilizer states as follows:
\begin{itemize}
    \item For the first 6 states, the spin directions of the two qubits are the same, i.e. $|\psi_\text{s} \>_1 = |+z\> \otimes |+z\>$, $|\psi_\text{s} \>_2 = |-z\> \otimes |-z\>$ etc.
    \item For the next 6 states, the spin directions of the two qubits are opposite, i.e. $|\psi_\text{s} \>_7 = |+z\> \otimes |-z\>$ etc.
    \item The next 24 states are organized into 6 groups, each containing 4 states where the spin of the two qubits are orthogonal, and the cross products of the two spins are the same within each group. For example, $|\psi_\text{s} \>_{13} = |+x\> \otimes |+y\>$, $|\psi_\text{s} \>_{14} = |+y\> \otimes |-x\>$, $\cdots$, $|\psi_\text{s} \>_{17} = |+x\> \otimes |-y\>$ etc.
\end{itemize}

\begin{table}[htbp]
    \centering
    \begin{tabular}{c|rrrrrrrrrrrrrrrrrrrr}
    \hline
       Order \# & 1 & 2 & 3 & 4 & 5 & 6 & 7 & 8 & 9 & 10 & 11 & 12 & 13 & 14 & 15 & 16 & 17 & 18 & 19 & 20  \\
       \hline
       $c_1$ & $1$ &  $0$ &  $0$ &  $0$ &  $1$ &  $1$ &  $1$ &  $1$ &  $1$ &  $1$ &  $1$ &  $1$ &  $1$ &  $1$ &  $1$ &  $1$ &  $1$ &  $1$ &  $1$ &  $1$ \\ \hline 
       $c_2$ & $0$ &  $0$ &  $1$ &  $0$ &  $1$ &  $-1$ &  $-1$ &  $1$ &  $i$ &  $-i$ &  $-i$ &  $i$ &  $i$ &  $-1$ &  $-i$ &  $1$ &  $-i$ &  $-1$ &  $i$ &  $1$ \\ \hline
       $c_3$ &$0$ &  $0$ &  $0$ &  $1$ &  $1$ &  $-1$ &  $1$ &  $-1$ &  $i$ &  $-i$ &  $i$ &  $-i$ &  $1$ &  $i$ &  $-1$ &  $-i$ &  $1$ &  $-i$ &  $-1$ &  $i$ \\ \hline
       $c_4$ & $0$ &  $1$ &  $0$ &  $0$ &  $1$ &  $1$ &  $-1$ &  $-1$ &  $-1$ &  $-1$ &  $1$ &  $1$ &  $i$ &  $-i$ &  $i$ &  $-i$ &  $-i$ &  $i$ &  $-i$ &  $i$ \\
\hline
       Order \# & 21 & 22 & 23 & 24 & 25 & 26 & 27 & 28 & 29 & 30 & 31 & 32 & 33 & 34 & 35 & 36 & 37 & 38 & 39 & 40 \\
       \hline
       $c_1$ & $1$ & $0$ & $0$ & $1$ & $1$ & $0$ & $0$ & $1$ & $1$ & $0$ & $0$ & $1$ & $1$ & $0$ & $0$ & $1$ & $1$ & $1$ & $1$ & $1$ \\ \hline
       $c_2$ & $-i$ & $1$ & $0$ & $0$ & $i$ & $1$ & $0$ & $0$ & $1$ & $1$ & $0$ & $0$ & $-1$ & $1$ & $0$ & $0$ & $0$ & $0$ & $0$ & $0$ \\ \hline
       $c_3$ & $0$ & $0$ & $1$ & $i$ & $0$ & $0$ & $1$ & $-i$ & $0$ & $0$ & $1$ & $-1$ & $0$ & $0$ & $1$ & $1$ & $0$ & $0$ & $0$ & $0$ \\ \hline
       $c_4$ & $0$ & $-i$ & $i$ & $0$ & $0$ & $i$ & $-i$ & $0$ & $0$ & $1$ & $-1$ & $0$ & $0$ & $-1$ & $1$ & $0$ & $1$ & $-1$ & $i$ & $-i$ \\
\hline
       Order \# & 41 & 42 & 43 & 44 & 45 & 46 & 47 & 48 & 49 & 50 & 51 & 52 & 53 & 54 & 55 & 56 & 57 & 58 & 59 & 60 \\
       \hline
       $c_1$ & $0$ & $0$ & $0$ & $0$ & $1$ & $1$ & $1$ & $1$ & $1$ & $1$ & $1$ & $1$ & $1$ & $1$ & $1$ & $1$ & $1$ & $1$ & $1$ & $1$ \\ \hline
       $c_2$ & $1$ & $1$ & $1$ & $1$ & $-1$ & $-1$ & $1$ & $1$ & $-i$ & $i$ & $-1$ & $-i$ & $-1$ & $i$ & $-i$ & $i$ & $1$ & $1$ & $-i$ & $i$ \\ \hline
       $c_3$ & $1$ & $-1$ & $i$ & $-i$ & $-1$ & $1$ & $-1$ & $1$ & $-i$ & $i$ & $-i$ & $-1$ & $i$ & $-1$ & $1$ & $1$ & $-i$ & $i$ & $i$ & $-i$ \\ \hline
       $c_4$ & $0$ & $0$ & $0$ & $0$ & $-1$ & $1$ & $1$ & $-1$ & $1$ & $1$ & $-i$ & $-i$ & $i$ & $i$ & $i$ & $-i$ & $i$ & $-i$ & $-1$ & $-1$ \\
\hline
    \end{tabular}
    \caption{The 60 stabilizer states for the 2-qubit system. The first 36 are unentangled states, while the rest are of maximal entanglement. We show the coefficients $\{c_i\}$ in the state $|\psi\> = c_1 |\uaua\> + c_2 | \uada\> + c_3 |\daua\> + c_4 | \dada\>$, up to overall phases and a normalization factor of $1$, $1/2$ or $1/\sqrt{2}$.}
    \label{tab:stabs}
\end{table}

\bibliography{ref}

\end{document}